\documentclass[12pt]{article}
\begin{document}
\title{A generalized Ginsparg-Wilson relation} 
\author{C.~D.~Fosco$^a$
  \,and\, M.~Teper$^b$
  \\
  {\normalsize\it $^a$Centro At\'omico Bariloche, 8400 Bariloche,
    Argentina}\\
  {\normalsize\it $^b$University of Oxford, Department of Physics,
    Theoretical Physics,}
  \\
  {\normalsize\it 1Keble Road, Oxford OX1 3NP, United Kingdom}}
\date{\today} \maketitle
\begin{abstract}
\noindent  We  show   that,  under certain   general  assumptions, any
sensible lattice Dirac operator satisfies a generalized form of the
Ginsparg-Wilson relation (GWR).  Those assumptions, on the other hand,
are mostly dictated by large momentum behaviour considerations.  We
also show that all the desirable properties often deduced from the
standard GWR hold true of the general case as well; hence one has,
in fact, more freedom to modify the form of the lattice Dirac
operator, without spoiling its nice properties.  Our construction, a
generalized Ginsparg-Wilson relation (GGWR), is satisfied by some
known proposals for the lattice Dirac operator. We discuss some of
these examples, and also present a derivation of the GGWR in terms of
a renormalization group transformation with a blocking which is not
diagonal in momentum space, but nevertheless commutes with the Dirac
operator.
\end{abstract}
\bigskip
\section{Introduction}
The construction of chirally symmetric theories on the lattice has
been the subject of renewed interest during the last years, because of
some interesting developments.  Among these, the overlap
formalism~\cite{neuber1}, based on an earlier idea by
Kaplan~\cite{kaplan}, has been shown to be a quite satisfactory
approach to the study of chirality-sensitive perturbative and
non-perturbative phenomena on the lattice.

On the other hand, the Nielsen-Ninomiya theorem~\cite{nini} tells us
that a strictly chirally symmetric theory on the lattice necessarily
breaks some nice features one would expect the theory to have, like
proper continuum limit for the fermion propagators, absence of
doublers, and locality. However, there is a compromise solution, which
amounts to breaking chiral symmetry, but in such a way that most of
the interesting properties are preserved.  Than can be achieved if the
Dirac operator $D$ satisfies the Ginsparg-Wilson relation~\cite{gins}
\begin{equation}\label{gwr}
D \gamma_5 + \gamma_5 D \,=\, a D \gamma_5 D \;.
\end{equation}
which  can be   thought of as   a  mild  deformation of  the  `strong'
anticommutation relation,
\begin{equation}\label{acr}
D_c \gamma_5 + \gamma_5 D_c \,=\, 0 \,,
\end{equation}
corresponding to a chirally symmetric operator $D_c$.

An important consequence of (\ref{gwr}), is that the fermionic action
defined in terms of such Dirac operator~\footnote{We shall follow here
  the conventions of~\cite{lusch}.}
\begin{equation}\label{defsf}
S_F \,=\, a^4 \sum_x {\bar\psi} D \psi \;,
\end{equation}
is invariant under the global symmetry transformation~\cite{lusch}
\begin{eqnarray}\label{gwt}
\delta \psi  &=&  i  \,\xi \,\gamma_5 (    1 - \frac{1}{2}  a  D) \psi
\nonumber\\   \delta {\bar\psi}  &=&  i   \,\xi \,  {\bar\psi}  ( 1  -
\frac{1}{2} a D)\gamma_5 \,,
\end{eqnarray}
where $\xi$ is a (real) infinitesimal parameter.  The finite-$\xi$
version of (\ref{gwt}) is, of course,
\begin{eqnarray}\label{fgwt}
\psi   &\to&   \exp     [ i   \,\xi      \,\gamma_5  (1-\frac{1}{2}  a
D)]\psi\nonumber\\   {\bar\psi}     &\to& {\bar\psi}   \exp[i      \xi
(1-\frac{1}{2} a D)\gamma_5]\,.
\end{eqnarray}
Thus the fermionic action has, on the lattice, a symmetry which in the
naive continuum limit ($a \to 0$) coincides with the usual chiral
transformation. Note, however, that the lattice symmetry transformation
itself is quite non-local.

Besides allowing for the existence of this symmetry, it has been
emphasized~\cite{chiu1,chiu2} that the general solution of (\ref{gwr})
can be written as
\begin{equation}\label{sgwr}
D\,=\, D_c  (  1 + \frac{1}{2} a  D_c)^{-1}  \,=\,( 1 + \frac{1}{2}  a
D_c)^{-1} D_c \;,
\end{equation} 
$D_c$ being a lattice Dirac operator satisfying (\ref{acr}).  This
implies the important property that $D$ and $D_c$ have the same
zero modes. Indeed, it has been shown that
\begin{equation}
{\rm index} (D) \;=\; {\rm index} (D_c) \;,
\end{equation}
and this is 
the reason why (\ref{sgwr}), as a mapping from $D_c$ to $D$, has been
called~\cite{chiu3} a `topological transformation'.

Thus the problem reduces to finding a suitable (non-local) operator $D_c$,
satisfying the relation (\ref{acr}).  It should also have the proper
index (the topological transformation cannot change it).  Inserting
that operator into (\ref{sgwr}) yields $D$, which is no longer
chirally symmetric ($\{D,\gamma_5\}\neq 0$), but which, for
a suitable  $D_c$, will be local.  This
could be summarized by saying that the non-locality of the chirally
symmetric $D_c$ is traded for the locality of $D$, at the price of
introducing non-local chiral symmetry transformations.

In this paper, we shall argue that the GWR is the simplest form of the
more general GGWR, and that the important properties often associated
to the GWR are also valid for the generalized relation.  Moreover (see
section~\ref{examples} below), this generalization appears quite
naturally in some examples.  The different realizations of the GGWR
will be shown to correspond to different ways of fixing the UV
behaviour of the Dirac operator for large eigenvalues.  As for the
case of the GWR, its generalization also requires the existence of a
suitable $D_c$, while the relation between $D$ and $D_c$ will be
different to~(\ref{sgwr}).  This introduces new parameters and
different ways of adjusting the locality properties of $D$.

The structure of this paper is as follows: in section~\ref{general},
we derive the generalization of the standard GWR, discussing the
nature of the different constraints one has to impose, and their
consequences. Then the general solution of the GGWR is presented, in
terms of the solution of a (c-number) equation.  We also discuss the
meaning of the GGWR in the context of renormalization group
transformations with a non-trivial averaging function.

Section~\ref{topo} deals with the `topological' properties of the 
lattice Dirac operator, such as its index theorems.  Finally, in
section~\ref{examples}, we consider some known proposals for lattice
Dirac operators, showing that they satisfy a GGWR.

\section{Generalizing the GWR}\label{general}

We begin by observing that, for the standard GWR case, the finite
transformations (\ref{fgwt}) that leave the fermionic action
(\ref{defsf}) invariant are {\em non-local}, because they include an
exponential of the (local) operator $D$, while they become local in
the infinitesimal case. It is rather strange to have a {\em global\/}
symmetry that makes such a radical distinction between finite and
infinitesimal transformations.  A natural, more symmetric situation,
would be to consider transformations that are not necessarily local,
even at the infinitesimal level:
\begin{eqnarray}\label{ggwt}
\delta \psi  &=& i  \,\xi\, \gamma_5  f(a  D) \psi \nonumber\\  \delta
{\bar\psi} &=& i \,\xi\, {\bar\psi} f(a D)\gamma_5 \,,
\end{eqnarray}
where $f$ is a real function with the proper continuum limit behaviour
\begin{equation}\label{limit}
x \,\to\, 0 \;\;\Rightarrow\;\; f(x) \, \to \, 1 \;,
\end{equation}
a condition that must be satisfied if we want (\ref{ggwt}) to be the
lattice version of the continuum chiral symmetry transformations. This
is the first of a series of constraints on $f$; the following ones
will appear, as we shall see, by relating $f$ to the actual form of
the lattice Dirac operator.  We note that relation (\ref{gwt}) is, of
course, a particular case of (\ref{ggwt}), i.e., \mbox{$f(x) = 1 -
  \frac{1}{2} x$}. Before proceeding, we remark that everything we
discuss in this section can be applied also to a theory in the
continuum, if the basic assumptions are true for the system under
consideration. Of course, one must then replace $a$ by $\Lambda^{-1}$,
where $\Lambda$ is the UV cutoff.

Of course not every $D$ will make (\ref{defsf}) invariant under
(\ref{ggwt}).  To see this, we {\em impose\/} (\ref{ggwt}) to be a
symmetry of (\ref{defsf}), and find that this requires $D$ to satisfy
\begin{equation}\label{ggwr}
D \,\gamma_5 \,f(a D) \,+\, f(a D)\, \gamma_5 \,D \,=\, 0 \;,
\end{equation}
which is the GGWR~\footnote{It is trivial to check that it yields the
  usual GWR if $f$ is assumed to be linear.}.

To begin with our study of (\ref{ggwr}), we first look for its
solutions, a problem which, except for the particular case of a linear
$f$, would seem to involve solving a non-linear operatorial equation.
This is fortunately not so, as we shall now see.  We first left and
right multiply both sides of (\ref{ggwr}) by \mbox{$\frac{1}{f(a
    D)}$}, obtaining
\begin{equation}\label{ir}
\frac{1}{f(a D)} \,D \,\gamma_5  \,+\, \gamma_5 \,\frac{1}{f(a D)} \,D
\,=\,0 \;,
\end{equation}
where we used the fact that $f(a D)$ commutes with $D$. It is evident
from (\ref{ir}) that the general solution is
\begin{equation}\label{imp}
\frac{1}{f(a D)} D \,=\, D_c
\end{equation} 
with $D_c$ as in the case of the usual GWR\@.  Now the non-linear
operatorial equation (\ref{imp}) yields an implicit definition for $D$
in terms of $D_c$.  The problem is greatly simplified by noting that
from (\ref{imp}), and from the fact that $f$ commutes with $D$, 
one deduces
\begin{equation}
 D\,=\,f(a D) D_c \,=\, D_c f(a D)
\end{equation}
namely, $f(a D)$ commutes with $D_c$.  Hence, also $D$ commutes with
$D_c$,
\begin{equation}\label{comm}
D D_c \,=\, f(a D) D_c^2 \,=\, D_c f(a D) D_c \,=\, D_c D \,.
\end{equation}
As $D$ and $D_c$ commute, we see that the problem of solving equation
(\ref{imp}) explicitly for $D$ in terms of $D_c$ reduces to that of
finding the (numerical) function $g$ such that $x \,=\, g(y)$ solves:
\begin{equation}\label{nleq}
\frac{x}{f(x)}\,=\, y \;.
\end{equation}
For such a solution to exist around a given point $x$, a necessary
condition is
\begin{equation}
\frac{d}{dx} \frac{x}{f(x)} \; \neq \; 0
\end{equation}
or
\begin{equation}
f(x) \; \neq \; C \, x
\end{equation}
with $C$ a real constant. We may then be sure that such a $g$ exists,
at least near to $x = 0$, because of condition (\ref{limit}).  To
address different $x$'s, we shall now study in more detail the
relation between $D$ and $D_c$.  It is clear that, when such a $g$
exists, then we can relate $D$ and $D_c$ by:
\begin{equation}\label{sggwr}
a D \,=\, g (a D_c) \;,
\end{equation}
which is the generalization of (\ref{sgwr}).  We note that, also
because of condition (\ref{limit}),
\begin{equation}\label{zero}
x \sim 0 \;\; \Rightarrow \;\; g(y) \sim y
\end{equation}
and thus $D \,\sim\, D_c$ for $a D \sim 0$.  This means that both $D$
and $D_c$ have approximately the same low-lying modes~\footnote{To
  work with Hermitian operators, one can of course multiply
  (\ref{sggwr}) to the left by $\gamma_5$.}.  This is true in
particular of the zero modes (see section~\ref{topo} below).

Let us now consider the large-$x$ regime.  We note that the fermionic
determinant (in an $A$ background), ${\mathcal Z}(A)$, corresponding
to the action (\ref{defsf}) can be written as
\begin{equation}
{\mathcal Z}(A)\;=\; \det  a D(A)  \;=\; \det  \left[ a \gamma_5  D(A)
\right] \;=\; \Pi_n \, ( a \lambda_n )
\end{equation}
where $\lambda_n$ are the eigenvalues of $\gamma_5 D$, which we order
according to their increasing moduli.  The $a$'s were introduced for
the sake of the normalization, and the $\gamma_5$ for algebraic
simplicity (note that $D$ and $\gamma_5 D$ have identical
determinants).

Because of (\ref{sggwr}), we may also write
\begin{equation}
{\mathcal Z}(A)\;=\; \Pi_n \,g(a \lambda_n^c)
\end{equation}
where $\lambda_n^c$ are the eigenvalues of $\gamma_5 D_c$.  The large
eigenvalue (large $n$, because of our ordering) regime is
characterized by $a \lambda_n^c >> 1$.  A necessary condition for the
determinant to be finite is
\begin{equation}
\lim_{n \to \infty} g(a \lambda_n^c) \;=\; 1
\end{equation}
since this is required for the convergence of the infinite product. 
Namely,
\begin{equation}\label{limit1}
y \; \to \; \infty \;\;\Rightarrow \;\; g(y) \;\to\; 1 \;.
\end{equation}
Condition (\ref{limit1}), when combined with (\ref{nleq}), yields
another important constraint on $f$:
\begin{equation}\label{limit2}
x \; \to \; 1 \;\;\Rightarrow \;\; f(x) \;\to\; 0 \;.
\end{equation}
The most immediate consequence of (\ref{limit2}) is that the symmetry
transformation (\ref{ggwt}) is trivialized at the large eigenvalues,
namely, it becomes the identity transformation for modes of the order
of the cutoff. More explicitly,
\begin{equation}\label{trivial}
\gamma_5 \, f (a D) \, \phi \; = \; 0
\end{equation}
if $\phi$ is an eigenvector of $\gamma_5 D$, with an eigenvalue of the
order of the cutoff.

Considering both (\ref{limit}) and (\ref{limit2}), we may conclude
that the generalized transformations (\ref{ggwt}) have the property of
behaving as the usual chiral transformations for low eigenvalues
(large distances), and disappearing for large eigenvalues (short
distances).  The non-local nature of the transformation stems from the
fact that it interpolates between these two regimes.

As a final remark on the general solution of (\ref{ggwr}), we check
that for the linear-$f$ case (GWR), we have \mbox{$f(x)\,=\, 1 -
  \frac{1}{2} x$}, and \mbox{$g(y)\,=\,\frac{y}{1+\frac{y}{2}}$}, thus
\begin{equation}
D \,=\, D_c ( 1 + \frac{1}{2} a D_c)^{-1} \;,
\end{equation}
as it should be.
\noindent Contrary to what happens  for the linear-$f$ case, not every
choice of $f(x)$ will allow us to find $D$ in terms of $D_c$ exactly.
However, there are important non-linear cases where such a solution
can indeed be found, as discussed in section~\ref{examples}.

It is convenient to adopt the convention that $\gamma_5 D_c$ is
Hermitian. Together with (\ref{sggwr}), this implies that $\gamma_5 D$
is also Hermitian, as can be easily verified.

Let us now study the GGWR from the point of view of the symmetry that
remains after performing a renormalization group transformation, which
is the approach followed in the original derivation of the
GWR~\cite{gins}.  To that end, we start by considering ${\mathcal
  Z}[A]$, the partition function for massless Dirac fermions in the
presence of an external gauge field:
\begin{equation}\label{eq:defza}
  {\mathcal Z}[A] \;=\; \int {\mathcal D}{\bar\psi} {\mathcal D}\psi \, 
  e^{- S_F[{\bar\psi},\psi;A]}
\end{equation}
where $S_F$ denotes the fermionic action. Since we want to deal with
both the continuum and lattice cases, we use the notation:
\begin{equation}
  S_F[{\bar\psi},\psi;A]\;=\; \int_{x,y} {\bar\psi}_x D_c(x,y)\psi_y 
\end{equation}
where the integral should be interpreted as a sum, when considering
the lattice case.  $D_c(x,y)$ is the chirally symmetric Dirac
operator. Then, following~\cite{gins}, we introduce a blocking
transformation to obtain an effective action for the new fermion
variables ${\bar \chi}, \chi$:
\begin{equation}\label{eq:rgt}
e^{-S_{eff}[{\bar\chi},\chi;A]}\;=\; \frac{1}{\mathcal N} \int
{\mathcal D}{\bar\psi} {\mathcal D}\psi \,
\times e^{-S_F[{\bar\psi},\psi;A] - {\mathcal T} [{\bar\chi},\chi;{\bar\psi},\psi;A]} \;,
\end{equation}
where ${\mathcal N}$ is a normalization constant and ${\mathcal T}$
defines the blocking.  The general form of this functional, for the
non-interacting fermion case is
\begin{equation}\label{eq:blockf}
  {\mathcal T} [{\bar\chi},\chi;{\bar\psi},\psi;A]\;=\; 
\int_{x,y}({\bar\psi}_x-{\bar\chi}_x) \alpha(x,y) (\psi_y -\chi_y)
\end{equation}
where alpha determines, of course, the properties of the
transformation. When it is a local function, diagonal in Dirac and
flavour space, the usual GWR emerges~\cite{gins}.  This can be done
following the original derivation in the reference above, or by
explicitly evaluating the functional integral over the original
fields, and comparing with (\ref{sgwr}). Performing the functional
integral over the fermionic fields, we obtain
\begin{equation}\label{eq:seff}
  S_{eff}[{\bar\chi},\chi;A]\;=\; \int_{x,y} {\bar\chi}_x D(x,y)\chi_y 
\end{equation}
with
\begin{equation} \label{eq:srgt}
  D\;=\; \frac{D_c}{1 \,+\, \alpha^{-1} D_c}\;.
\end{equation}
The simplest choice corresponds to taking a constant $\alpha$, which
by dimensional reasons has to be proportional to the cutoff, $\alpha
\propto \Lambda \propto 1/a$. In fact, relation follows from the
choice $\alpha \,=\, \displaystyle{\frac{2}{a}}$.

We now show that the GGWR may be obtained from a blocking
transformation which is non diagonal in Dirac space. We note that
$\alpha$ has to commute with $D_c$, otherwise, more
symmetries would be broken. As this has to happen regardless of the
particular background field configuration, $\alpha$ can only be a
function of $D_c$:
\begin{equation}
\alpha \;=\;  \frac{2}{a \, \beta[a D_c ]} \;,
\end{equation}
where $\beta$ is a dimensionless real function. In this case, we have
of course
\begin{equation}\label{eq:srgt1}
D\;=\; \frac{D_c}{1 \,+\,\frac{a}{2} 
\beta[a D_c ] D_c} \;.
\end{equation}
This allows we to consider the effect of iterating the blocking
transformation.  Denoting by $D^{(0)}$ the initial operator
$D_c$, and by $D^{(n)}$ the one that results
from applying the transformation $n$ times, we see that
\begin{equation}\label{eq:iterated}
  D^{(n)}\;=\;  \frac{D_c}{1 \,+\, 
\frac{n a}{2}\beta[a D_c ] D_c}\;.
\end{equation}

That the operator $D$ in (\ref{eq:srgt1}) will satisfy a GGWR is evident
from the fact that the relation between $D$ and $D_c$ implies, in the
notation of section~\ref{general},
\begin{equation}
x \; =\; g (y) \;=\; \frac{y}{1 + \frac{1}{2} \beta(y) y} \;,
\end{equation}
and this implies a non linear $f(x)$, except for a trivial function
$\beta$.

\section{Topological properties}\label{topo}
Let  us first check the  property  of `topological invariance' of  the
transformation (\ref{sggwr}),  namely, whether it  preserves the index
of $D_c$ or not. To that end, let $\phi_0^{\pm}$ be a chiral zero mode
of $D_c$:
\begin{equation}
D_c \, \phi_0^{\pm} \;=\; 0
\end{equation} 
where $\pm$ denotes the chirality of $\phi_0^{\pm}$
\begin{equation}
(\frac{1\pm \gamma_5}{2}) \phi_0^{\pm} \,=\, \phi_0^{\pm} \;.
\end{equation}
Then,
\begin{equation}
D \phi_0^{\pm} \,=\, \frac{1}{a} g(a D_c) \phi_0^{\pm} \;,
\end{equation}
which does, indeed, yield
\begin{equation}
D \phi_0^{\pm} \,=\,0
\end{equation}
We are using property (\ref{zero}) for the regular behaviour of $g(x)$
near to the origin. As $g(x) \sim  x$ for $x  \to 0$, then the inverse
of $g$ also behaves like $\sim x$ around  $x = 0$. This guarantees the
validity of  the reciprocal proposition: a zero  mode of $D$ is also a
zero mode of $D_c$. Thus  (\ref{sggwr}) is topologically invariant, as
expected from the fact that the lowest modes  of $D$ and $D_c$ are not
distorted by that transformation.

Let us now study the  index theorems on the  lattice.  We shall follow
the approach developed in~\cite{fuji0,fuji},  adapting it to the GGWR\@.
As a first  check,  we verify  that the  global  version of  the index
theorem,  obtained   from   the  anomalous   Jacobian   for a   chiral
transformation, holds true  for  the GGWR\@.  Following~\cite{fuji0}, we
have to consider the trace
\begin{equation}
{\rm Tr} \left[\gamma_5 f( a D) \right]
\end{equation} 
which  corresponds   to  half  of   the   Jacobian  factor  for    the
transformation (\ref{ggwt}).  Evaluating the trace in the basis of the
$\phi_n$, eigenvectors of the Hermitian  operator $\gamma_5 D$, we see
that
$$
{\rm  Tr}  \left[\gamma_5    f(  a D)   \right]\;=\;  \sum_n
\phi_n^\dagger \, \gamma_5 f( a D) \, \phi_n 
$$
\begin{equation}\label{ind0}
=\;  \sum_{\lambda_n  =   0}  \phi_n^\dagger \gamma_5    \phi_n  \,+\,
\sum_{\lambda_n \neq 0} \phi_n^\dagger \gamma_5 f( a D) \phi_n
\end{equation}
as a consequence of (\ref{limit}). To deal  with the sum over non-zero
modes,  we  left  multiply  relation  (\ref{ggwr})  by  $\gamma_5$ and
sandwich it with an arbitrary non-zero mode $\phi_n$, obtaining
\begin{equation}\label{nonzeig}
\lambda_n \, \phi_n^\dagger \gamma_5 f(a D) \phi_n \;=\;- \lambda_n \,
\phi_n^\dagger \gamma_5 f(a D) \phi_n
\end{equation}
where we applied the property:
\begin{equation}
\phi_n^\dagger \gamma_5 D \;=\; \lambda_n \phi_n^\dagger \,.
\end{equation}
As  $\lambda_n  \neq 0$,   then \mbox{$\phi_n^\dagger \gamma_5  f(a D)
\phi_n = 0$}.  Thus we immediately deduce that
\begin{equation}\label{ind01}
\sum_{\lambda_n \neq 0} \phi_n^\dagger \gamma_5 f( a D) \phi_n \,=\, 0
\;.
\end{equation}
Inserting this result back into (\ref{ind0}), we obtain a result which
is identical to the one in~\cite{fuji0}, namely
\begin{equation}\label{ind1}
\sum_n \phi_n^\dagger \gamma_5  f( a D) \phi_n \,=\,\sum_{\lambda_n  =
0}   \phi_n^\dagger   \gamma_5 \phi_n \,=\,   n_+  -   n_-  \,=\, {\rm
index}(D).
\end{equation}

The above theorem, (\ref{ind01}) and (\ref{ind1}), means that
\begin{equation}
{\rm Tr}\, \Gamma_5 \;=\; n_+ - n_- \;,
\end{equation}
where $\Gamma_5 \,=\,   \gamma_5 \,f(a D)$.  Because of  (\ref{ggwr}),
this operator anticommutes with the Hermitian operator ${\cal D} \,=\,
\gamma_5 D$:
\begin{equation}\label{indx}
{\cal D} \, \Gamma_5 \,+\, \Gamma_5 \, {\cal D} \;=\; 0 \;,
\end{equation}
which is identical to the relation found by Fujikawa in~\cite{fuji}.

As  for the  interpretation of  the ${\rm  Tr}\gamma_5 =  0$ relation,
discussed in~\cite{fuji} for the case of the  GWR, we follow a
parallel approach. We should therefore find the `highest states', namely,
zero  modes  of $\Gamma_5$.   These  are again responsible  for  the
cancellation of the contribution coming from the  zero modes of $D$ to
the  trace of $\gamma_5$. Denoting  by $\phi$ those highest
states, we should then have
\begin{equation}
\gamma_5 \, f(a D) \phi \;=\; 0 \;,
\end{equation}
but this is precisely  the relation we found in (\ref{trivial}),
namely, those vectors do exist {\em as a consequence of the good large
momentum behaviour of $D$}.  As  for  their number (degeneracy),  it
should be the  same as for  the zero modes of   ${\cal D}$, since  the
degeneracy depends on  the symmetries (and  they are the same for both
operators).

\section{Examples}\label{examples}
In 't Hooft's approach~\cite{thooft}, the  fermionic determinant is 
defined {\em  in the continuum\/} by
\begin{equation}\label{defdet}
\det D \;=\; \Pi_i \; \left[ \det (D_c + \Lambda_i) \right]^{e_i}
\end{equation}
where
\begin{equation}
D_c \,=\,   \not \!\!   D   \,=\, \gamma_\mu D_\mu  \,\,,\,\, D_\mu  =
\partial_\mu + i A_\mu \,. 
\end{equation}
Here $D_c$ stands  for `continuum' Dirac  operator, and indeed it also
verifies the  chirality  condition (\ref{acr})  satisfied  by the {\em
lattice\/}  operator   $D_c$.    To      render $\det   D$      finite
(non-perturbatively),  the gauge field $A_\mu$  is  defined through an
extrapolation mechanism  from  the usual link variables~\cite{thooft},
and   besides,    Pauli-Villars    regulators    are  introduced    in
(\ref{defdet}), verifying the usual conditions:
\begin{equation}
\sum_i e_i \log \Lambda_i \;\equiv \log \Lambda \;\;,\;\; 
\sum_i e_i \Lambda^n \log \Lambda_i \;\equiv \; 0 \;. 
\end{equation}
Here $n$  runs  from  $0$   through $4$  if   one  is dealing with   a
four-dimensional theory.  Of course, this range  will be different for
different numbers of dimensions.   In $4$ dimensions,  three regulator
fields are sufficient, while in $2$ dimensions only one.  We shall now
see that $D$ will satisfy a GWR when there is  only one regulator, and
a GGWR when three (or more) regulators are required.

The simplest case  corresponds to  using   just one regulator   field,
which, in order to improve the UV behaviour of the bare Dirac operator
has to have bosonic statistics. This is frequently  stated in terms of
a regulated action $S_F^R$
\begin{equation}\label{sfr1}
S_F^R \,=\,\int d^d x \left[ {\bar \psi}(x) \not \!\! D \psi (x) \,+\,
{\bar\phi}(x)(\not \!\! D + \Lambda)\phi (x) \right]
\end{equation} 
where $\phi$ is the bosonic regulator field~\footnote{This regularized
action was also introduced in the GWR context by Fujikawa~\cite{fuji},
although not    in  order  to   show that  it   satisfies   the GWR.}.
Equivalently, if one integrates out  the regulator, the result is  the
equivalent action (for which we use the same notation)
\begin{equation}\label{sfr2}
S_F^R \,=\,\int d^d x \, {\bar \psi}(x) D_R \psi (x)
\end{equation} 
where
\begin{equation}\label{spv}
D_R \,=\, D_c (\frac {1}{\Lambda} D_c + 1)^{-1} \;.
\end{equation}
This is not the usual way of writing the  action, since it is formally
non-local,   and   the       version   (\ref{sfr1})   is     sometimes
preferred. However, (\ref{sfr2}) is the  more convenient form to study
the GWR: the operator $D_R$, as it is evident by comparing (\ref{spv})
with (\ref{sgwr}), will verify the relation
\begin{equation}
D_R  \gamma_5 + \gamma_5 D_R \,=\,  \frac{2}{\Lambda} D_R \gamma_5 D_R
\,
\end{equation} 
i.e., a GWR with $a=\frac{2}{\Lambda}$.   We can of course borrow some
of the already known properties of the GWR to  this case. For example,
the regularized    action    (\ref{sfr2})  is invariant     under  the
transformation:
\begin{eqnarray}\label{ctrans}
\delta \psi &=& i \,\xi \, \gamma_5 \, ( 1 - \frac{D_R}{\Lambda}) \psi
\,=\,   i \,\xi \,  \gamma_5  \,  \frac{\Lambda}{D_c  + \Lambda}  \psi
\nonumber\\  \delta {\bar\psi}    &=& i  \,\xi   \,{\bar\psi}  (  1  -
\frac{D_R}{\Lambda})   \gamma_5  \,=\, i     \,\xi \,   {\bar\psi}  \,
\frac{\Lambda}{D_c + \Lambda}\gamma_5
\end{eqnarray}
where  we have expressed  the transformations  both   in terms  of the
regularized and unregularized Dirac operators.

This symmetry for the continuum theory has  consequences for the study
of anomalous  Ward identities,  from  the point  of view of  anomalous
(Fujikawa) Jacobians.      Let  us   recall   that, in      the  usual
setting~\cite{fuji1},  one starts    from  the   functional   integral
representation for the unregularized fermionic determinant $D_c$,
\begin{equation}
\det D_c \;=\; \int {\cal D}{\bar\psi} {\cal D}\psi \, \exp\{-S_F \}\,
\end{equation}
and then performs a local axial transformation for the fermions:
\begin{eqnarray}
\delta  \psi (x) \;=\; i  \,  \xi (x)\,  \gamma_5 \psi (x) \nonumber\\
\delta  {\bar\psi} (x) \;=\; i \,  \xi (x)  \, {\bar \psi}(x) \gamma_5
\;.
\end{eqnarray} 
This produces a variation in the action:
\begin{equation}
\delta S_F \;=\;-i \int d x \, \xi (x) \, \partial_\mu J_{\mu 5}
\end{equation}
where   \mbox{$J_{\mu  5}={\bar\psi}\gamma_\mu\gamma_5\psi$}, and  the
measure acquires an anomalous Jacobian $J$:
\begin{equation}
{\cal D}{\bar\psi} {\cal D}\psi  \to \, J  \, {\cal D}{\bar\psi} {\cal
D}\psi \;.
\end{equation}
$J$ is ill defined,
\begin{equation}
J  \;=\; \exp[ -2 i  {\cal A} ]\;\;,\;\; {\cal  A}\;=\;  {\rm Tr}( \xi
\gamma_5 )\;.
\end{equation}
A regulator $\rho$ is then introduced to give meaning to the trace,
\begin{equation}
{\cal   A}\;=\;   \int    dx  \,    \xi (x) {\rm      tr}  [  \gamma_5
\rho(\frac{\not\!\!D^2}{\Lambda^2}) ]
\end{equation}
where $\rho(0)=1$,   and $\rho$ vanishes rapidly  for  $x \to \infty$.
Evaluating the regulated Jacobian  and  taking then the $\Lambda   \to
\infty$ limit yields the well known result:
\begin{equation}
{\cal         A}   \;=\;    \int         dx    \,    \xi    (x)     \;
\frac{1}{32\pi^2}\epsilon^{\mu\nu\alpha\beta}F_{\mu\nu}F_{\alpha\beta}
\end{equation}
which leads to the anomalous Ward identity
\begin{equation}
\partial_\mu    \langle   J_{\mu   5}\rangle  \;=\;  \frac{1}{16\pi^2}
\epsilon^{\mu\nu\alpha\beta}F_{\mu\nu}F_{\alpha\beta}\;.
\end{equation}

On the other hand, as we know  about the existence of  the GWR and its
symmetry for the  continuum  theory, can  start from the   regularized
action  (\ref{sfr1}), perform the  transformation (\ref{ctrans}),  and
then  evaluate the anomalous Jacobian,   {\em which does  not need any
extra  regularization}, since  its  exponent is proportional  to the
functional trace:
\begin{equation}
{\rm Tr} \left[\gamma_5 (1-\frac{D_R}{\Lambda})\right]
\end{equation}
and moreover, in terms of $D_c=\not \!\! D$,
\begin{equation}\label{fin}
{\rm Tr}  \left[\gamma_5 (1-\frac{D_R}{\Lambda})\right] \,=\, {\rm Tr}
\left[\gamma_5 \frac{1}{1 + \frac{\not  \! D}{\Lambda}}\right] =  {\rm
Tr} \left[\gamma_5 \frac{1}{1 + \frac{\not \!  D^2}{\Lambda^2}}\right]
\;.
\end{equation}
And  (\ref{fin}) is one  of the general  expressions which lead to the
proper result for  the chiral anomaly  in the continuum, when $\Lambda
\to \infty$. The variation of the action, just produces the divergence
of a regularized current.  We thus again obtain  the usual result, but
note that the procedure has the advantage of involving regular objects
at all the steps in the calculation.

It   is  well known  that more   general Pauli-Villars regularizations
amount to replacing the simple equation (\ref{spv}) by:
\begin{equation}\label{genpv}
\frac{D_R}{\Lambda} \,=\, {\cal R}(\frac{D_c}{\Lambda}) \;,
\end{equation}
where ${\cal  R}$ denotes a rational  function,  and $\Lambda$  is the
cutoff.   The  function  ${\cal  R}$  will depend  also   on  a set of
dimensionless constants.

It is obvious that   (\ref{genpv}) plays a similar role to (\ref{sggwr}),
and in  order to find the corresponding  $f$ (and hence  the GGWR), we
have to find the solution to
\begin{equation}
y = \frac{x}{f(x)} \;.
\end{equation}
For the particular case of a four dimensional theory, and if one wants
to have convergence for all the diagrams contributing to the fermionic
determinant (vacuum diagrams factorized) the rational function is
\begin{equation}
{\cal R}(x) \;=\; y \, \frac{y + \sqrt{2}}{(y + 1)^2} \;.
\end{equation}
This expression was  obtained by integrating  the  regulator fields in
the corresponding regularized action:
\begin{equation}
S_F^R \;=\; \int d^4 x \sum_{s=0}^3 {\bar \psi}_s (D_c + M_s) \psi_s
\end{equation}
where $\psi_0$ and $\psi_1$  are fermionic fields, while  $\psi_{2,3}$
are bosonic.  The masses are: $M_0=0$,  $M_1^2= 2 \Lambda^2$, $M_2^2 =
M_3^2 = \Lambda^2$.

For this particular case, the relation can, indeed, be inverted to
obtain $y$ as a function of $x$, and so we can find the form of the
function $f(x)$, which determines the GGWR and its associated global
symmetry.  The answer is:
\begin{equation}
f(x)\;=\;\frac{{\sqrt{2}}\,\left(   1    -  x  \right)    \,x}  {-1  +
{\sqrt{2}}\,x + {\sqrt{1 - 2 ({\sqrt{2}}-1)\,x}}}
\end{equation}

We have seen that the continuum Dirac operator plays here the same
role as the lattice operator $D_c$, whose drawback is its
non-locality.  This is also true of 't Hooft's method, since the
continuum Dirac operator, when viewed from the lattice point of view,
is a non-local object.

As another example of a Dirac operator which verifies a GGWR, we point
our Slavnov's proposal~\cite{slavnov}. It amounts to using, on the
lattice, a SLAC derivative lattice Dirac operator (which has no
doublers and is non-local), supplemented by Pauli-Villars regulator
fields. These regulators, with cutoffs of the order of $a^{-1}$, have
the effect of taming the lattice artifacts introduced by the SLAC
derivative operator. Indeed, it has been shown that more than one
regulator field may be necessary to have a reasonable behaviour, even
in two dimensions~\cite{slavnov}. When three regulators are used, we
are in the situation of the previously discussed example, where a GGWR
relation is verified.

\underline{Acknowledgements}:

C.D.F. acknowledges Profs. H.~Neuberger and T-W.~Chiu for useful comments.
This work commenced when one of the authors (M.~T.) was visiting
the Centro At\'omico Bariloche. He is grateful to the Centro
for its hospitality, and to the Fundaci\'on Antorchas - British Council exchange 
program for funding his visit.
C.~D.~F. was supported by CONICET, and ANPCyT (Argentina).

\end{document}